\documentclass[twocolumn,amsmath,amssymb,pra]{revtex4-2}

\usepackage{graphicx}
\usepackage{color}
\usepackage{dcolumn}
\usepackage{amsmath}
\usepackage{CJK}
\usepackage{subfigure}
\usepackage{sidecap}
\usepackage{dsfont}

\usepackage{mathrsfs}
\usepackage{amsfonts,amssymb}
\usepackage{indentfirst}
\usepackage{bm}

\usepackage{relsize}

\usepackage{physics}
\usepackage{soul}

\usepackage[dvipsnames]{xcolor}

\usepackage[colorlinks,citecolor=blue]{hyperref}

\newcommand{\be}{\begin{equation}}
\newcommand{\ee}{\end{equation}}
\newcommand{\bey}{\begin{eqnarray}}
\newcommand{\eey}{\end{eqnarray}}
\newcommand{\bw}{\begin{widetext}}
\newcommand{\ew}{\end{widetext}}

\newcommand{\ra}{\rangle}
\newcommand{\la}{\langle}

\newcommand{\ba}{\begin{array}}
\newcommand{\ea}{\end{array}}
\newcommand{\bi}{\begin{itemize}}
\newcommand{\ei}{\end{itemize}}
\newcommand{\bem}{\begin{enumerate}}
\newcommand{\eem}{\end{enumerate}}

\newcommand{\jamil}{\textcolor{blue}}

\begin{document}

\title{Characterizing the Lipkin-Meshkov-Glick excited spectrum through the quantum coherence spectrum
}

\author{Qian Wang\footnote{Electronic address: qwang@zjnu.edu.cn}}

\affiliation{Department of Physics, Zhejiang Normal University, Jinhua 321004, China}
\affiliation{CAMTP-Center for Applied Mathematics and Theoretical Physics, University of Maribor, Mladinska 3, SI-2000
Maribor, Slovenia}

\author{Jamil Khalouf-Rivera}
\affiliation{Departamento de Ciencias Integradas y Centro de Estudios Avanzados en F\'{i}sica,
Matem\'{a}ticas y Computaci\'{o}n, Universidad de Huelva, Huelva 21071, Spain}

\author{Francisco P\'{e}rez-Bernal\footnote{Electronic address: curropb@uhu.es}}

\affiliation{Departamento de Ciencias Integradas y Centro de Estudios Avanzados en F\'{i}sica,
Matem\'{a}ticas y Computaci\'{o}n, Universidad de Huelva, Huelva 21071, Spain}
\affiliation{Instituto Carlos I de F\'{i}sica Te\'{o}rica y Computacional, Universidad de Granada, Granada 18071, Spain}

\begin{abstract}
Excited-state quantum phase transitions extend the quantum phase transition concept beyond the ground state and offer insights into the complex behavior of quantum systems. In the present work, we assess the use of the multiple quantum coherence spectrum as a valid tool to probe excited state quantum phase transitions within the framework of the Lipkin-Meshkov-Glick model. The time dependence and the long-time average of the multiple quantum coherence spectrum reveal the intricate dynamics that stems from the existence of singularities in the excited spectrum of a quantum many-body system. 
\end{abstract}

\date{\today}

\maketitle

\section{introduction} 

Quantum phase transitions, also known as ground-state 
quantum phase transitions (GSQPTs), 
are non-thermal phase transitions characterized by 
abrupt changes in the ground state of a quantum system 
once one or more Hamiltonian control parameters reach certain critical values~\cite{Sachdev1999,Carr2010}. 
This concept was enlarged to encompass excited states, which  leads to
the definition of excited-state quantum phase transitions (ESQPTs) 
\cite{Cejnar2006,  Cejnar2008,Caprio2008}. 
For a recent review on the subject see Ref.~\cite{Cejnar2021}. 
Unlike GSQPTs, that convey non-analytical changes in the system ground-state properties, 
 ESQPTs are associated with nonanaliticities in the level flow and density 
 of states (DOS) of the system in its mean-field or 
 large system size limit~\cite{Stransky2016, Cejnar2021}. 
 ESQPTs are found in various many-body quantum systems, 
 e.g.~the Lipkin-Meshkov-Glick (LMG) model
\cite{Leyvraz2007,Engelhardt2015,Sindelka2017,Santos2016,Nader2021,Gamito2022}, 
the Dicke and Rabi models 
\cite{Magnani2014,Fernandez2011,Brandes2013,Puebla2013,PerezFP2017,Puebla2016}, 
spinor Bose-Einstein condensates \cite{Feldmann2021,Cabedo2021,Niu2024},  
ferromagnetic quantum gases \cite{Meyer2024},
the vibron model \cite{Larese2011, Larese2013,Jamil2021,Jamil2022, KRivera2019,KRJamil2022, KRivera2024}, 
periodically-driven systems \cite{Bastidas2014,Garcia2021},  the interacting boson model \cite{Macek2019,DongW2021}, 
the coupled top model \cite{WangP2021,Mondal2022}, 
as well as Kerr nonlinear oscillators \cite{WangS2020, Chavez2023, Prado2023, Iachello2023}.

The nonanalytical effects associated with ESQPTs have clear precursors at finite system sizes that modify the system structure and dynamics. The presence of a ESQPT is usually revealed in the closing of the gap between adjacent energy levels, the appearance of exceptional points, or extremal values of quantities such as the order parameter of the associated GSQPT, the participation ratio or the quantum fidelity susceptibility~\cite{Santos2015,Sindelka2017,Gamito2022,Jamil2022,KRJamil2022}. Thermal properties of systems with a finite number of degrees of freedom can also be strongly modified due to the existence of an ESQPT with anomalous thermal relaxation or the presence of a singularity in the microcanonical heat capacity in the vicinity of the ESQPT critical energy~\cite{Cejnar2017, Relaño2018}. Finally, ESQPTs can also have a strong impact on system dynamics as revealed in the study of the system under sudden quantum quenches, with strong effects on the survival probability of the initial state and a maximally enhanced decoherence, in the study of the time dependence of out-of-time order correlators (OTOCs), where the enhanced information scrambling associated with the unstable character of the ESQPT critical energy is revealed, and in peculiar features in non-equilibrium thermodynamics~\cite{Relano2008,Fernandez2011,Santos2016,Kloc2018,WangP2019b,Stransky2021,Kloc2021,Gamito2022II,Chavez2023,Puebla2015,WangH2017,Mzaouali2021}. Recently, ESQPT precursors have also been identified in dissipating systems, modeled by complex-valued Hamiltonians \cite{Kopylov2015,Rubio2022}.

Quantum coherence is one of the cornerstones of quantum theory and it plays an important role in
various quantum technologies \cite{Streltsov2017}.
For any given basis, a pure quantum state 
can be expressed using its density matrix, emphasizing the concept of probability distribution in the quantum formalism and removing the arbitrary overall phase that is implicit when expressing quantum states as vectors in a Hilbert space. The diagonal elements of the density matrix are known as \textit{populations} or zeroth-order \jamil{of the Multiple Quantum Coherence (MQC) spectrum} while non-diagonal elements are dubbed \textit{coherences} or $n$-th order \jamil{of the MQC spectrum}, with $n$ depending on the order of the diagonal in the density matrix. Non-zero coherences in a pure quantum state density matrix imply that the corresponding elements of the selected basis are in a quantum superposition, introducing quantum interference effects. Since its introduction in nuclear magnetic resonance studies \cite{Baum1985,Baum1986,keeler2011},  the MQC spectrum has proved a valuable tool  with different applications. It gives access to many-body quantum coherences \cite{Baum1985,Cappellaro2014,Macieszczak2019} and to the evolution of correlations \cite{Munowitz1987,Sanchez2014} and entanglement  \cite{Doronin2003,Furman2008,Feldman2012}.  Beyond the usual experimental access to QMC spectra in nuclear magnetic resonance studies, recent advances in cold atom physics (ion traps, ultracold atoms, or Bose-Einstein condensates) paved the way to the access to QMC for pure quantumn states at temperatures close to zero. In particular, G\"arttner \textit{et al.} measured the MQC spectrum using a trapped-ions setup~\cite{Garttner2017}, demonstrating the link between MQC spectra and OTOCs, and suggesting an experimentally feasible MQC protocol that provides access to mutiparticle entanglement trough quantum Fisher information~\cite{Garttner2018}.

The MQC spectrum is also a valid proble to detect GSQPTs in Hermitian \cite{Swan2020,Deng_2024} and
non-Hermitian systems \cite{Pires2021}, as well as the transition to localization in disordered systems~\cite{Alvarez2010,Alvarez2015,WeiKX2018}. In this article we investigate how the existence of an ESQPT in the LMG model modifies quantum coherence and whether the MQC spectrum is a valid probe for ESQPTs.

The LMG model was first introduced in nuclear physics to describe interacting fermions with a toy model, simple enough to be solved exactly, that could be used to validate different approximations~\cite{LIPKIN1965188,Meshkov1965199,Glick1965211}. Despite its simplicity, it revealed as a versatile model, with enough physical content to be used in other disciplines. It has proved a convenient platform for studies on quantum phase transitions
\cite{Caprio2008,Cejnar2021,Defenu2018,Johannes2018,Corps2022,
Botet1983,Dusuel2004,Leyvraz2005,Ribeiro2007,Titum2020}, 
quantum thermodynamics \cite{Campbell2016,MaYu2017,Hardal2018,Mzaouali2021,ZhangZ2022}, 
quantum metrology \cite{Salvatori2014,GuanQ2021,Garbe2022},
quantum control \cite{Caneva2008,Puebla2020,Abah2022}, 
or quantum information \cite{Vidal2007,Wilms2012,Chiara2018,BaoJ2020,HuM2021,Bao2021}. It has also been employed to explore the influence of unstable stationary points on the spreading 
of OTOCs \cite{Cameo2020, Xu2020}, 
the behavior of the complexity in a system with infinite
range interaction \cite{Kunal2022},
the existence of Floquet time-crystals in a 
system without disorder \cite{Russomanno2017}, and to link the critical phenomena of static Hamiltonians with the features observed in periodic quenched systems~\cite{saiz2024quantum}. 
A very positive aspect of this model is that it can be accomplished in a variety of experimental platforms~\cite{Albiez2005,Zibold2010,AFerreira2013, Jurcevic2014, Jurcevic2017, ZhangJ2017,Makhalov2019,Muniz2020,KaiXu2020,Cervia2021,Hlatshwayo2022,Li2023}. 

The LMG model Hamiltonian in the present paper undergoes 
a second order GSQPT and an associated ESQPT in the GSQPT broken-symmetry phase, which is
characterized by a logarithmic divergence of the DOS at the critical energy \cite{Stransky2016}.
Such divergence is a consequence of an unstable 
fixed point appearing in the dynamics of the classical limit of the model.
We show that the occurrence of the ESQPT leads to a drastic change in the MQC spectrum for
time-evolved and long-time averaged  states.
We also illustrate how to identify ESQPT signatures from the behavior 
of the MQC spectrum zero mode and we demonstrate 
that the  MQC spectrum width is a reliable ESQPT probe. 
Thus, the aim of the present paper is to provide a detailed exploration of the interplay between an
ESQPT and the MQC spectrum and to evince the 
usefulness of the MQC spectrum as an ESQPT detector in many-body quantum systems.

The article is structured as follows.
In Sec.~\ref{SecS}, we begin with a brief review to the concept of MQC spectrum.
In Sec.~\ref{ThrS}, we present the LMG model and its 
classical limit, making emphasis on the model
ESQPT main features.
Our main results are given in Sect.~\ref{ForS}.
On the first hand, in Subsect.~\ref{ForSa} we discuss 
the time dependence of the MQC spectrum of a state 
once the system undergoes a quantum quench. 
On the second hand, in Subsect.~\ref{ForSb}, 
we study how the MQC spectrum width is a valid ESQPT probe. 
We end the results section, in Subsect.~\ref{ForSc}, with a study of the MQC 
spectrum for the resulting long-time averaged state after the quench. 
Finally, we summarize our results and conclusions in Sec.~\ref{FivS}.

\section{Multiple quantum coherence protocol} 
\label{SecS}
The MQC protocol, initially introduced in nuclear magnetic resonance studies~\cite{Baum1985,Lacelle1991, Feldman2008}, 
provides information about multi-particle quantum coherences, 
the non-diagonal elements of the density matrix. Let's consider a Hermitian operator $\hat{ \mathcal{O}}$, with 
real eigenvalues $\nu_n$ and associated eigenstates $\ket{\nu_n}$. 
The density matrix of an arbitrary pure quantum state,  
$\rho$, can be expressed in the $\{\ket{\nu_n}\}$ basis as  
\be \label{Dcmrho}
   \rho=\sum_\ell\rho_\ell=\sum_\ell \sum_{\nu_n-\nu_m=\ell}\rho_{nm}|\nu_n\ra\la\nu_m|~,
\ee
with $\rho_{nm}=\la\nu_n|\rho|\nu_m\ra$ and where the index $\ell$ takes all possible values determined by the $\nu_n-\nu_m$ differences. 
Each $\rho_\ell$ block is known as  an $\ell$-coherence 
and it contains information about coherences 
between eigenstates of $\hat{\mathcal{O}}$ 
whose eigenvalues differ by $\ell$, $\nu_n-\nu_m=\ell$. 
The $\ell$-th multiple quantum intensity, \(I_\ell(\rho)\), is defined as the square of the Frobenius norm 
(also known as Hilbert-Schmidt norm or Schatten $2$-norm) of $\rho_\ell$~\cite{Garttner2018,Swan2020,Pires2021},
\be \label{DfMQI}I_\ell(\rho)\equiv(||\rho_\ell||_2)^2=\mathrm{Tr}[\rho_\ell^\dag\rho_\ell]~.
\ee
The set of all $I_\ell(\rho)$ intensities forms the $\rho$ state MQC spectrum. 
Note that the sum of the multiple quantum intensities  over all possible 
$\ell$ values is one if and only if $\rho$ is a pure quantum state. 
Apart from the information about the system coherences 
provided by the $\ell\neq 0$ intensities, the $\ell = 0$ intensity depends on the density matrix populations, the diagonal matrix elements of the density matrix.

As shown in Refs.~\cite{Garttner2017,Garttner2018, Lewis2019}, multiple quantum intensities can be experimentally accessed through a protocol defined in three steps. A given initial pure quantum state, $\hat \rho_i$, is evolved for a time $t$ under a nontrivial Hamiltonian resulting in $\hat \rho_t = \hat U(t)\hat \rho_i \hat U^\dagger(t)$, where $\hat U(t) = e^{-i\hat H t}$. This is followed by the application of the operator $\hat W(\phi) = e^{-i\hat{ \mathcal{O}}\phi}$, and a backward evolution of time $-t$ producing the final state $\hat \rho_f$. It can be shown that the fidelity $F_t(\phi) = \Tr{\hat\rho_i\hat \rho_f(\phi)}$, using cyclic permutations under the trace, is expressed as 
\begin{equation}\label{MQCFotoc}
 F_t(\phi) = \Tr{\hat\rho_i\hat \rho_f(\phi)}= \Tr{\hat\rho_t\hat \rho_t(\phi)} = \sum_\ell I_\ell e^{-i\ell\phi}~,  
\end{equation}
\noindent where $\hat \rho_t(\phi) = \hat W(\phi)\hat\rho_t\hat W^\dagger(\phi)$. From the previous equation it is clear the the Fourier transform of the fidelity with respect to $\phi$ gives access to the values of the different multiple quantum intensities. Another points developed in  Refs.~\cite{Garttner2017,Garttner2018, Lewis2019} is the connection of the MQC spectrum to an OTOC. Given two operators, $\hat V$ and $\hat W$, and an initial state, an OTOC measures the spread in time of the operator $\hat W$ through the expectation value of the commutator squared module $\expval{[\hat W_t, \hat V]^\dagger [\hat W_t \hat V]}$, where $\hat W_t = e^{i\hat H t} \hat W e^{-i\hat H t}$. This commutator has a term equals to a time disordered matrix element of the operators in the given initial state, $F_{\hat V,\hat W}(t) = \expval{\hat W^\dagger_t\hat V^\dagger \hat W_t \hat V}$. In the MQC case one can define $\hat V$ as the initial matrix density $\hat \rho_i$ and $\hat W$ as the  $\hat W(\phi)$ operator which for the case of small $\phi$ values can be shown to be a Fidelity OTOC (FOTOC) \cite{Lewis2019}.

OTOCs are sensitive to the sudden changes in the ground state associated with GSQPTs~\cite{Shen2017}, and the equivalence between MQC spectra and OTOCs indicates its suitability to detect and characterize such transitions, as demonstrated in Refs.~\cite{Swan2020,Deng_2024}. OTOCs have also proved useful in the detection and study of ESQPTs \cite{WangP2019b}. In the present article our aim is to characterize the LMG model ESQPT making use of the MQC spectrum and its second moment
\be \label{WMQCs}
\Sigma(\rho,\hat{\mathcal{O}})=\left[\sum_\ell\ell^2I_\ell(\rho)\right]^{1/2}~,
\ee
\noindent a quantity that recently has been proved to be equal to the quantum Fisher information for pure quantum states~\cite{Garttner2018}.

\begin{figure}
    \includegraphics[width=\columnwidth]{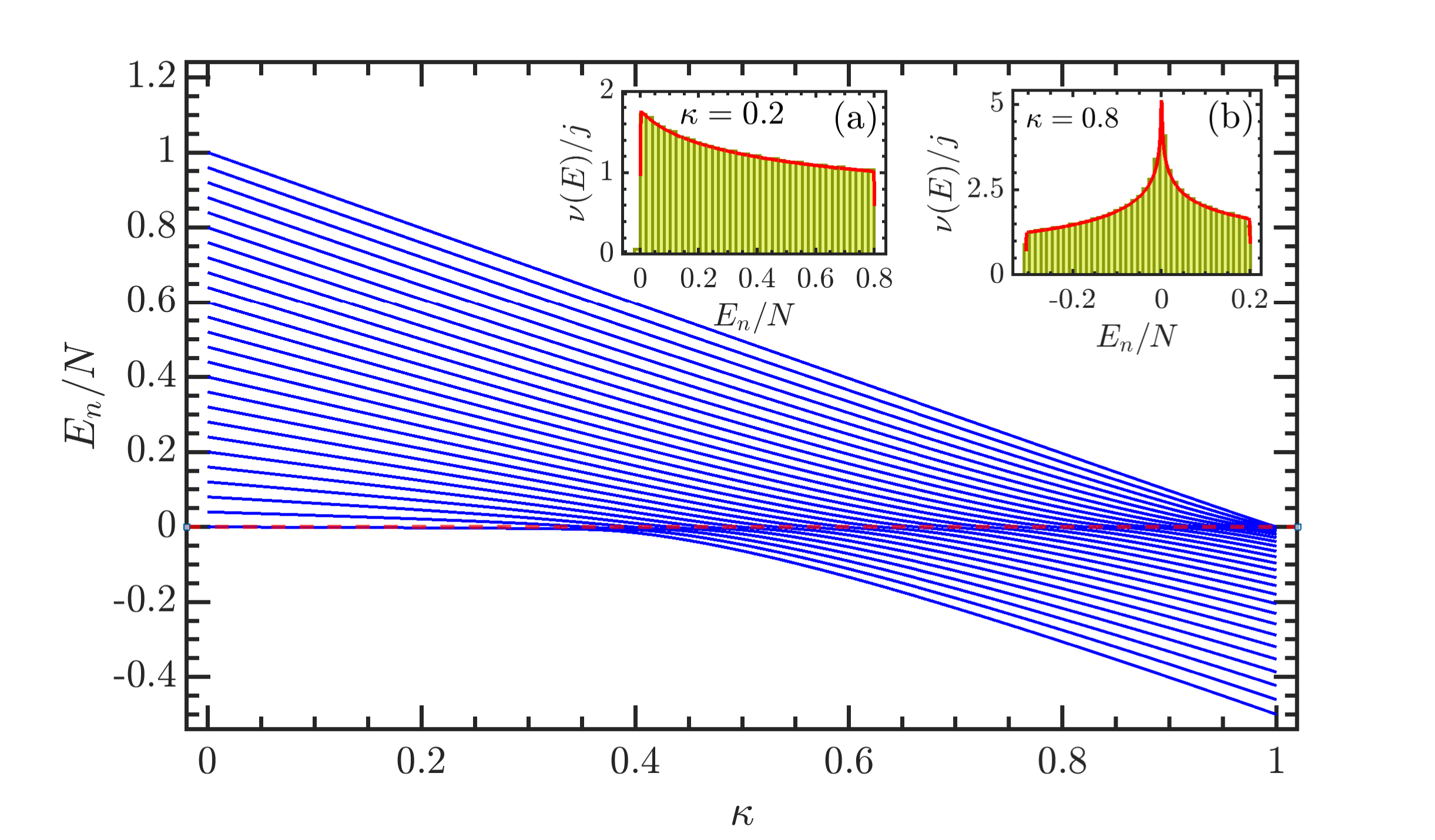}
    \caption{Normalized energy spectrum of Hamiltonian Eq.~(\ref{LMGH}) 
    as a function of the control parameter $\kappa$ with $j=N/2=25$.
    The dashed horizontal line marks the critical ESQPT energy, $E_c=0$.
    The scaled density of states is depicted in the two insets, for a $\kappa$ 
    value less than $\kappa_c$ (inset (a), $\kappa = 0.2$) and a value larger 
    than the critical one (inset (b), $\kappa = 0.8$) for a system size $N=5000$.
    The red solid line in each inset is the classical limit result, $\nu_c(E)$, 
    from Eq.~(\ref{CDOS}) (see main text).
    The ESQPT in the model is characterized by a divergence in the level density $\nu(E)$, as shown in the inset panel (b). All plotted quantities are dimensionless.}
    \label{Espectrum}  
\end{figure}

\section{The Lipkin-Meshkov-Glick model} \label{ThrS}

The LMG model has different mathematical realizations. Nowadays, the most frequently used one is a fully connected Ising model, a spin-$1/2$ chain with $N$ elements and all-to-all, infinite range, interactions. The Pauli matrix of the $i$-th spin along the $\alpha$-axis is $\hat{\sigma}_i^\alpha$ with $\alpha = x,y,z$ and $i=1, \ldots, N$. Considering collective quasispin operators, $\hat{J}_{\alpha}=\sum_{i=1}^N\hat{\sigma}_{i}^{\alpha}/2$, the spin chain Hamiltonian can be recast as a simpler collective Hamiltonian with a second-order GSQPT and an associated ESQPT
\begin{align} \label{LMGH}
   \hat H=\varepsilon\left[-\frac{2\kappa}{N}\hat{J}_x^2+(1-\kappa)\left(\hat{J}_z+\frac{N}{2}\right)\right]~,
\end{align}
with a control parameter $\kappa$ defined in the $0\leq\kappa\leq 1$ range. Here, and throughout the present work, we set $\hbar=1$ and the energy scale $\varepsilon=1$, making the Hamiltonian dimensionless. Hamiltonian Eq.~\eqref{LMGH} conserves the total number of particles, $N$. The use of collective quasispin operators implies a drastic reduction of the Hilbert space dimension, from a total dimension of $2^N$ in the spin chain case to $N+1$ if we only consider the totally symmetric representation spanned by Dicke states (symmetric states with all spins coupled to the maximum quasispin value, $j=N/2$). 
There exists an additional $\mathbb{Z}_2$ symmetry in Hamiltonian Eq.~\eqref{LMGH}, the parity symmetry $\hat{\Pi}=e^{i\pi(j+\hat{J}_z)}$, that divides the Hilbert space into even and odd non-mixing blocks. Hence, we can restrict our study to the even-parity block, reducing further the Hilbert space dimension to $(N- N \mod 2)/2+1$. The Hamiltonian (\ref{LMGH}) can be diagonalized 
in the Dicke basis, $|m\rangle$ defined as $J_z|m\rangle=m|m\rangle$, where $-j\leq m\leq j$, splitting the basis into two subsets according with the parity of the quantum number $m$.

The ground state of Hamiltonian (\ref{LMGH}) undergoes a second-order 
GSQPT at the control parameter critical value $\kappa_c=1/3$, which separates the symmetric phase ---also known as the normal or ferromagnetic phase--- 
($\kappa<\kappa_c$) from the broken-symmetry phase ---often called superradiant or paramagnetic phase--- ($\kappa>\kappa_c$).
The signatures of this GSQPT have been thoroughly investigated in literature
\cite{Botet1983,Dusuel2004,Dusuel2005,Leyvraz2005,Ribeiro2007,Romera2014,Campbell2016,
Titum2020,Ruiz2021}.
It is also known that Hamiltonian (\ref{LMGH}) exhibits an ESQPT for $\kappa>\kappa_c$, with a critical energy $E_c=0$~
\cite{Caprio2008,Perez2009,Santos2016,Cejnar2021,Gamito2022}.
In this work, we characterize the LMG ESQPT  
using an MQC protocol.

\begin{figure}
    \includegraphics[width=\columnwidth]{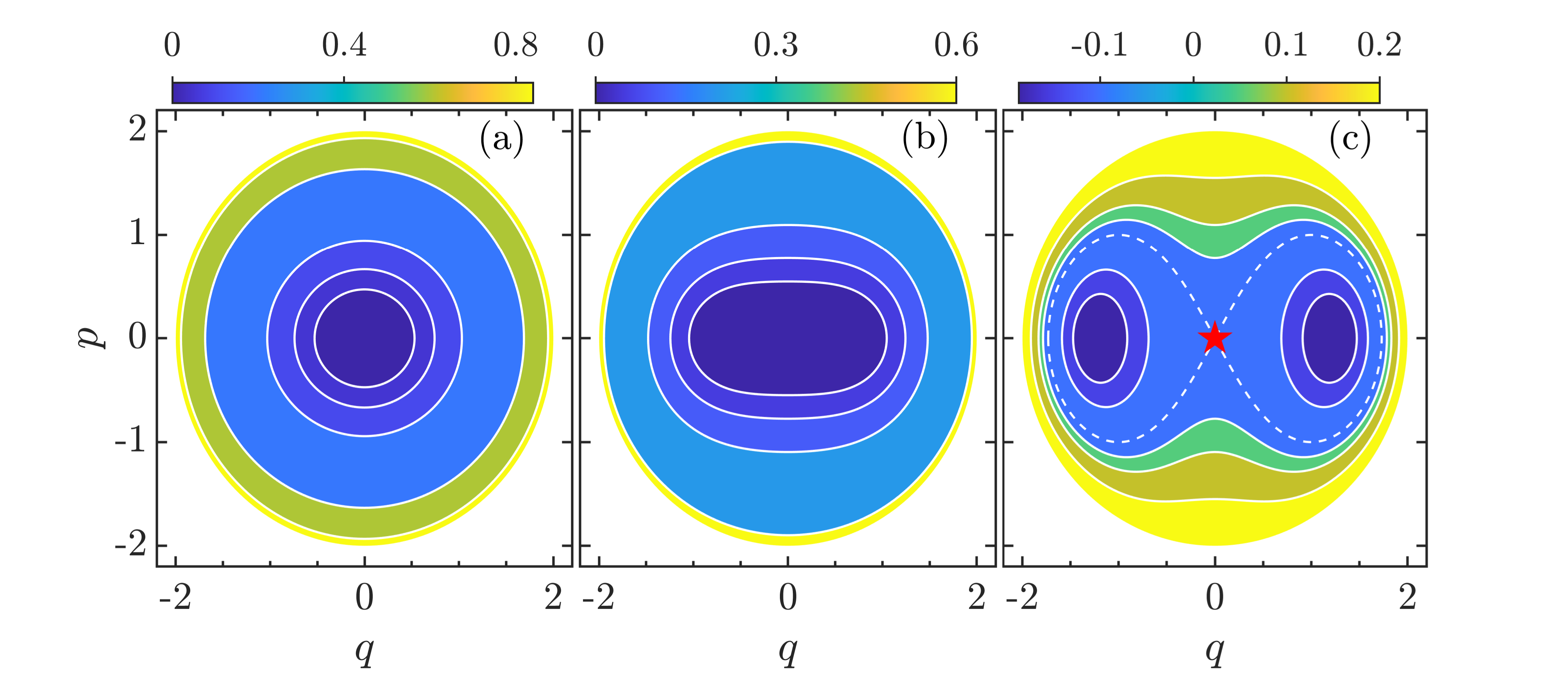}
    \caption{Energy surface contour plots for the classical 
    Hamiltonian (\ref{CLMG}) with control parameter values
    $\kappa=0.3\kappa_c$ (a), $\kappa=\kappa_c$ (b), and $\kappa=2\kappa_c$ (c). 
    The red star and the dashed line in panel (c) denote the saddle point and 
    its corresponding energy separatrix in phase space, respectively.
   This unstable critical point in the classical 
    limit of the LMG model is associated with the ESQPT in the quantum system.
    The critical control parameter value is $\kappa_c=1/3$.
    All quantities are dimensionless.} 
    \label{Egsurface}  
\end{figure}

ESQPTs are characterized by a nonanalytic density of states at a certain excitation energy, the ESQPT critical energy \cite{Caprio2008,Stransky2016, Cejnar2021}.
In Fig.~\ref{Espectrum}, we show the correlation energy diagram for Hamiltonian Eq.~\eqref{LMGH} as a function of the 
control parameter $\kappa$ for a system with size $N=50$. A high density of states line can be appreciated at the normalized energy $E_c/N=0$. In the insets of Fig.~\ref{Espectrum}, we show the density of states,
defined as $\nu(E)=\sum_n\delta(E-E_n)$,
 for $\kappa = 0.2<\kappa_c$ in inset (a) and $\kappa = 0.8>\kappa_c$ in inset (b). The green bars have been numerically computed for a system with a size $N=5000$, while the red-solid line is the result obtained in the system mean-field limit as we explain below. There is a clear contrast between the $\kappa <\kappa_c$ results, with no ESQPT, and the high density of levels around $E_c/N=0$ for $\kappa>\kappa_c$~\cite{Perez2009,Santos2016,Santos2015,Nader2021}, which develops into a logarithmic divergence at the classical limit \cite{Stransky2016,Cejnar2021}. 

\begin{figure*}
    \includegraphics[width=\textwidth]{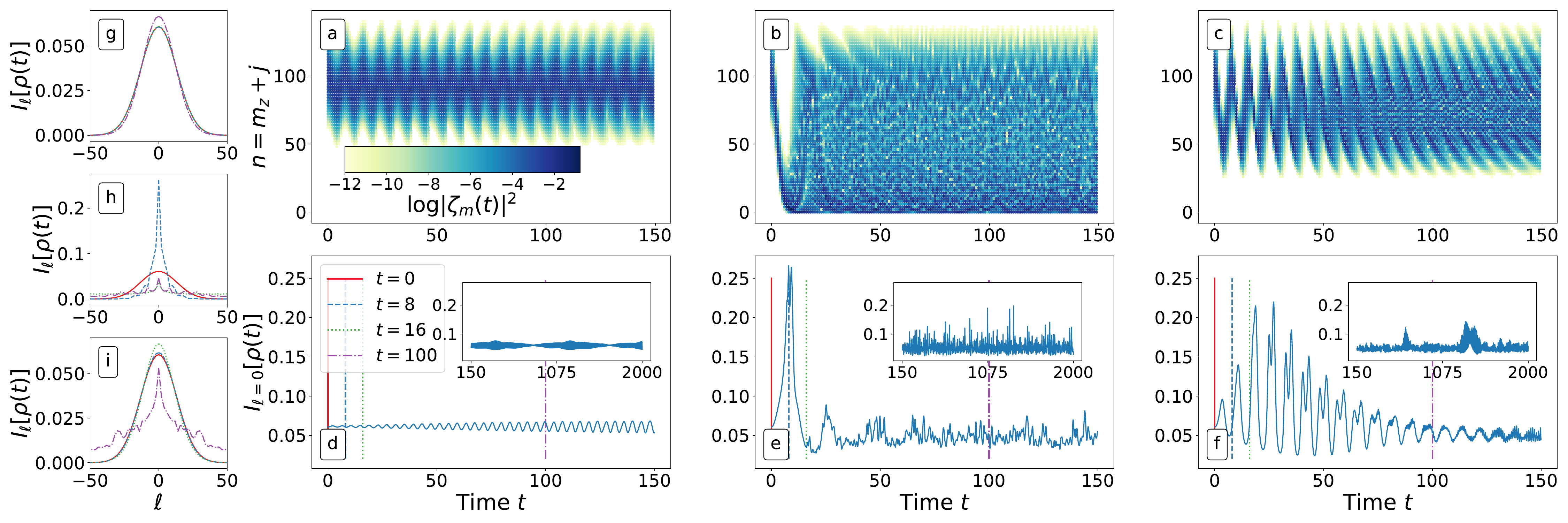}
    \caption{Panels (a)-(c): Heat maps displaying the coefficients $|\zeta_m(t)|^2$ 
    as a function of the $n=m_z+j$ quantum number  (ordinate) 
    and time (abscissas) for $\chi=0.2\chi_c$ (a), $\chi=\chi_c$ (b), and
    $\chi=2\chi_c$ (c) with $\kappa=1.5\kappa_c$ and a system size $N=400$. 
    Panels (d)-(f): Zeroth-order component of the MQC spectrum 
    $I_0[\rho(t)]$, defined in Eq.~(\ref{ZeroM}), 
    as a function of time $t$. Values of $\chi$, $\kappa$, and $N$ are the same than in panels (a)-(c). The insets display the same quantity in a larger time span. 
    The six plots share the abscissa axis range. 
    Panels (g)-(i): $I_\ell[\rho(t)]$ versus the differences $\ell$ for 
    the selected values of time (see panel (d) legend) highlighted 
    in panels (d)-(f) with vertical dashed lines, using the same color code and line style.
   The large spreading of $|\zeta_m(t)|^2$ and  its rapid decay to $n = 0$ value
    in panel (b) and the particular evolution of $I_0[\rho(t)]$
    in panel (e) are dynamical signatures of the LMG model ESQPT.
    Other parameter values: $\kappa_c=1/3$ 
    and $\chi_c$ given by Eq.~(\ref{CrQ}).}
    \label{QCspectrum}
\end{figure*}

The ESQPT in the LMG model can be understood considering the stationary points of 
the system classical limit, obtained for  $N\to\infty$~\cite{Cejnar2021}, using quasispin coherent states~\cite{Radcliffe1971,Zhang1990}
\begin{align}
  |\xi\rangle=\frac{1}{(1+|\xi|^2)^j}e^{\xi \hat{J}_+}|j,-j\rangle~,
\end{align}
where $\xi\in\mathbb{C}$ is a variational parameter and $\hat{J}_\pm=\hat{J}_x\pm i\hat{J}_y$ 
are quasispin raising and lowering operators.
The expectation value of the operators needed to study Hamiltonian Eq.~\eqref{LMGH} are~\cite{Radcliffe1971}
\begin{align}
\langle\xi|\hat{J}_+|\xi\rangle&=\frac{2j\xi^\ast}{1+|\xi|^2}~, \nonumber\\
  \langle\xi|\hat{J}_-|\xi\rangle&=\frac{2j\xi}{1+|\xi|^2}~,  \\  \langle\xi|\hat{J}_z|\xi\rangle&=j\left(\frac{|\xi|^2-1}{|\xi|^2+1}\right)~.  \notag
\end{align}
Taking this into account, the classical limit of Hamiltonian Eq.~\eqref{LMGH} is
\begin{align}\label{eq:primitiveH}
  \mathcal{H}_c(\xi)&=\frac{\langle\xi|\hat{H}|\xi\rangle}{N}  \notag  \\
     &=-\frac{\kappa(\xi+\xi^\ast)^2}{2(1+|\xi|^2)^2}+(1-\kappa)\frac{|\xi|^2}{1+|\xi|^2}~.
\end{align}
The complex parameter $\xi$ can be mapped into classical canonical variables $(p,q)$, considering the change of variables
\be
  \xi=\frac{q+ip}{\sqrt{4-(p^2+q^2)}}~,
\ee
and Eq.~\eqref{eq:primitiveH} transforms to the classical limit of the LMG Hamiltonian \eqref{LMGH}
\begin{align}\label{CLMG}
  \mathcal{H}_c(p,q)=-\frac{\kappa q^2}{8}(4-p^2-q^2)+\frac{1-\kappa}{4}(p^2+q^2)~.
\end{align}

The $(q,p)_s$ stationary points of $\mathcal{H}_c(p,q)$ are the solutions of 
$\nabla\mathcal{H}_c(p,q)=0$
\begin{align}
   (q,p)_s=
   \left\{
     \begin{aligned}
       &(0,0)  \\
       &\left(\pm\sqrt{\frac{3\kappa-1}{\kappa}}, 0\right) \quad  \text{for} \quad \kappa>\kappa_c
     \end{aligned}
    \right.~.
\end{align}
Note that for  $\kappa\le\kappa_c$ there is a single stationary point, a minimum at the origin, $(q,p)_s=(0,0)$. For control parameter values $\kappa >\kappa_c$, 
the stationary point in the origin is a saddle point and two 
identical minima appear at zero momentum, $(q,p)_s= (\pm\sqrt{\frac{3\kappa-1}{\kappa}}, 0)$.
The ground-state energy functional in the mean-field limit 
can be obtained by calculating the value of the classical 
Hamiltonian Eq.~\eqref{CLMG} evaluated in the corresponding global minimum
\begin{align}
  \mathcal{E}_{gs}(\kappa)=
  \left\{
   \begin{aligned}
     &0 \quad  \text{for}\quad \kappa\leq\kappa_c~,   \\
     &-\frac{(3\kappa-1)^2}{8\kappa} \quad \text{for}\quad \kappa>\kappa_c~.
   \end{aligned}
   \right.\label{gsefunc}
\end{align}
It can be easily shown that the second derivative of the energy functional Eq.~\eqref{gsefunc} with respect to the control parameter is discontinuous when evaluated at the critical value of the control parameter, as expected in a second-order GSQPT~\cite{Romera2014}.

Fig.~\ref{Egsurface} panels depict energy contour plots of the  
classical limit of the LMG model in Eq.~\eqref{CLMG} for
three different values of $\kappa$.
In Fig.~\ref{Egsurface}(a), we observe that, as mentioned above, 
for $\kappa<\kappa_c$ the energy surface has
a global minimum at the origin $(0,0)$.
At $\kappa=\kappa_c$ ---Fig.~\ref{Egsurface}(b)---, the system 
exhibits a flatter minimum (quartic order) that splits up 
into two symmetric wells for values of $\kappa>\kappa_c$ ---Fig.~\ref{Egsurface}(c).

For a fixed value of the control parameter, the available phase space volume depends on the energy, 
\begin{align}\label{CDOS}
  \nu(E)=\frac{1}{2\pi}\iint\delta[E-\mathcal{H}_c(p,q)]dpdq~.
\end{align}
Note that $\nu(E)$ is the smooth component in the Gutzwiller 
trace formula \cite{Gutzwiller2013}, identified as the semiclassical approximation to the quantum
density of states \cite{Caprio2008,Nader2021,Magnani2014}.

Following the approach outlined in Refs.~\cite{Nader2021,Magnani2014}, we 
plot $\nu(E)$ for two different values of $\kappa$ 
in the insets of Fig.~\ref{Espectrum} using solid-red lines.
The agreement between $\nu(E)$ and the numerical results (green bars) computed for a system with $N=5000$ is excellent.
Furthermore, in the broken-symmetry phase, the DOS exhibits a logarithmic divergence at the normalized critical energy $E_c/N=0$ due to the saddle point of the classical energy surface ---see red star in Fig.~\ref{Egsurface}(c)---~\cite{Caprio2008,Nader2021}.

\begin{figure}            
\includegraphics[width=\columnwidth]{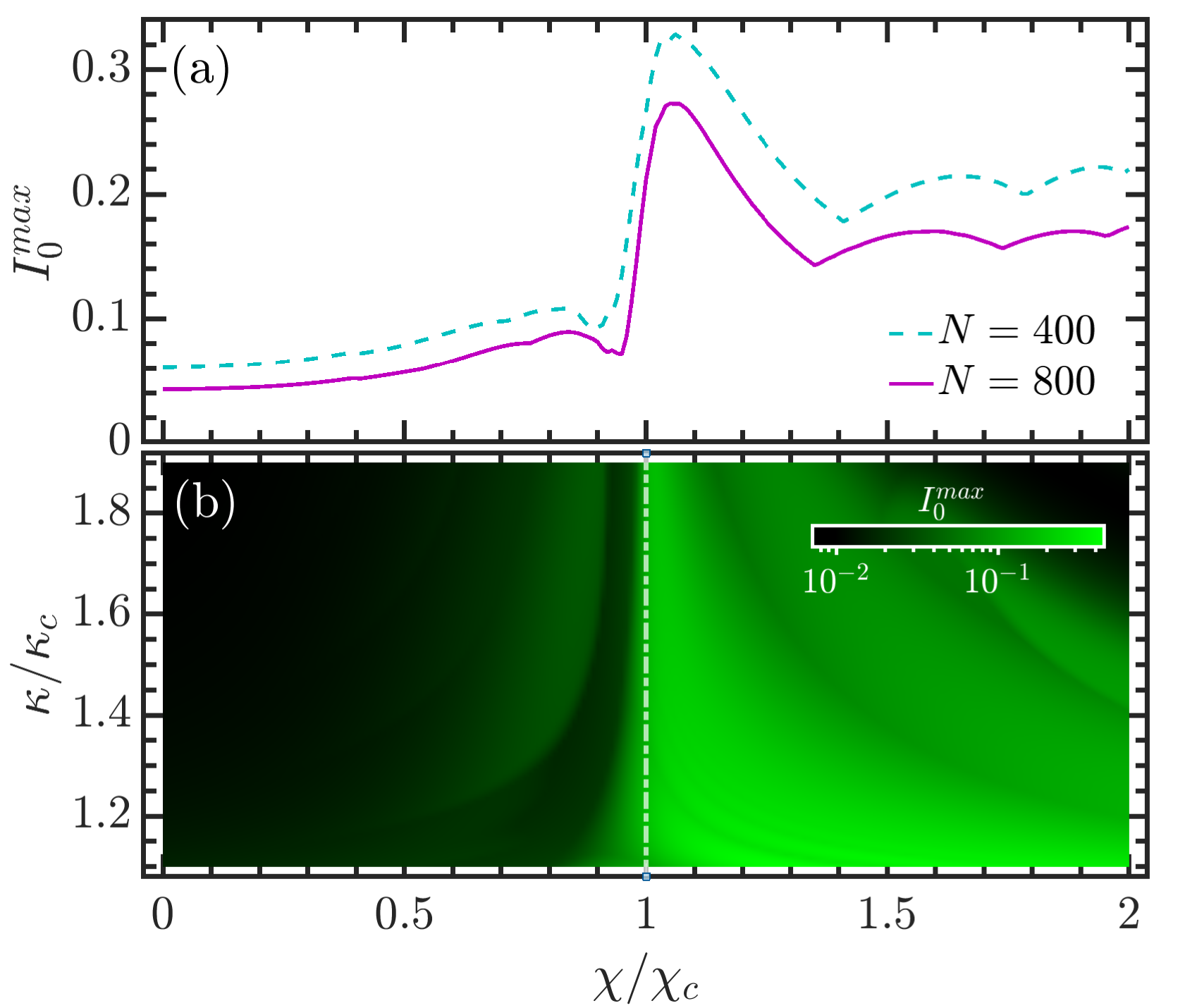}
    \caption{Panel (a): Maximum zero component of the 
    MQC spectrum, $I_0^{max}$ (see Eq.~(\ref{MaxI})), 
    as a function of $\chi/\chi_c$ for a system with $\kappa=1.5\kappa_c$, a time interval $t\in[0,\tau = 30]$,  
    and system sizes N=400 (green-dashed line) and $800$ (red-solid line).
    The rapid growth of $I_0^{max}$ near $\chi/\chi_c=1$ indicates
    the presence of ESQPT.
    Panel (b):  $I_0^{max}$ heat map as a function of $\chi/\chi_c$ and $\kappa/\kappa_c$ 
    for a system size $N=800$. 
    The vertical white dot-dashed line indicates the critical line $\chi/\chi_c = 1$.
    Parameter values: $\kappa_c=1/3$ 
    and $\chi_c$ given by Eq.~(\ref{CrQ}).} 
    \label{Zero}  
\end{figure}

\section{Results} \label{ForS}
To demonstrate the convenience of the MQC spectrum as a probe of the LMG model ESQPT 
in Hamiltonian Eq.~\eqref{LMGH}, we consider the following  sudden-quench protocol.
Initially, the system is prepared in a $\ket{\psi_0}$ state that is the even parity ground state of an LMG Hamiltonian $\hat{H}_0$, with a $\kappa$ control parameter value such that the system is in the broken-symmetry phase ($\kappa_c<\kappa<1$). At $t=0$,  a $\hat{J}_x^2$ interaction with strength $-2\chi/N$ is added to the Hamiltonian, giving rise to a non-trivial time-evolution of the initial state governed by Hamiltonian $\hat{H}_1=\hat{H}_0-(2\chi/N)\hat{J}_x^2$. The value of the quench parameter, $\chi$, allows us to bring the system to different excited-state phases. The critical value of the quench parameter, $\chi_c$, brings the initial state to the ESQPT critical energy,  $E_c/N=0$ and its value can be obtained from the 
classical approach using the tangent method~\cite{Fernandez2011,Gamito2022II}
\be \label{CrQ}
  \chi_c=-\frac{\kappa(3\kappa-1)}{\kappa+1}~,
\ee
with $\kappa_c<\kappa<1$. 
We would like to emphasize that
the ESQPT for the LMG model Hamiltonian (\ref{LMGH}) only exists in
the broken-symmetry phase, for control parameter values in the range $\kappa_c<\kappa<1$. As we pay heed to the ESQPT imprints in the MQC spectrum, the present study is restricted to the broken-symmetry phase.

\begin{figure}      
    \includegraphics[width=\columnwidth]{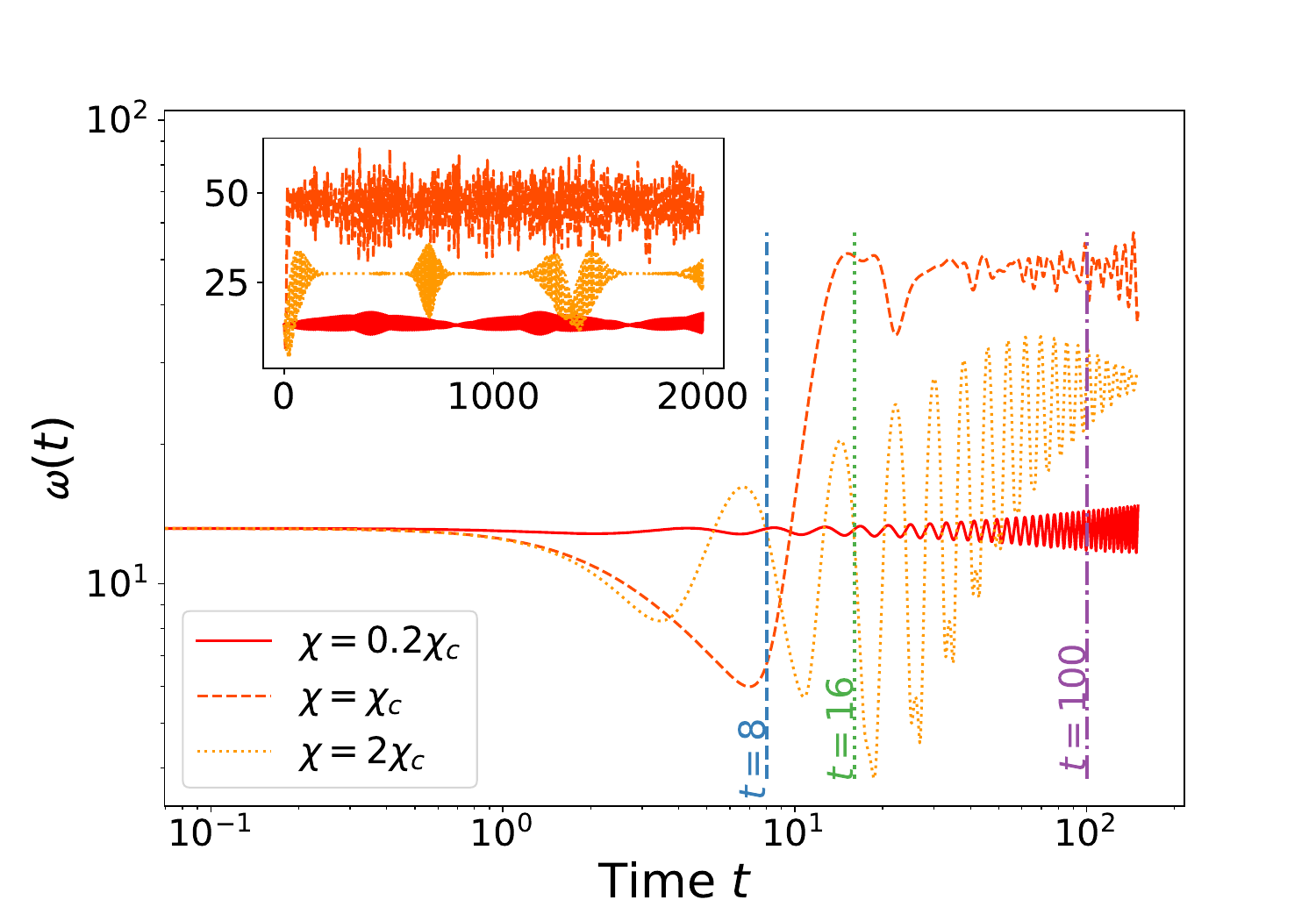}
    \caption{Time evolution of the MQC spectrum 
    width, $w(t)$ (see Eq.~(\ref{Dwidth})), in a log-log scale. 
    From dark to light orange: $\chi/\chi_c=0.2$ (full line), 
    $\chi/\chi_c=1$ (dashed line), and $\chi/\chi_c=2$ (dotted line),
    for a system with $\kappa=1.5\kappa_c$ and size $N=400$.
    The abscissa axis of the main panel shares its time 
    scale with Figs.~\ref{QCspectrum}(d)-(f). 
    In the inset, the same quantity is depicted versus time 
    using linear axes and for a longer time span.
    The rapid growth in the dynamics of $w(t)$ for the critical case
    can be trace back to the LMG ESQPT.
    Other parameter values: $\kappa_c=1/3$ 
    and $\chi_c$ is defined in Eq.~(\ref{CrQ}).
    }
    \label{Timewidth}
\end{figure}

\subsection{Quenched state MQC spectrum} \label{ForSa}

We first calculate the MQC spectrum of the state at a time $t$ after the quench
\be
  \rho(t)=|\psi_t\rangle\langle\psi_t|=e^{-iH_1t}\rho(0) e^{iH_1t},
\ee
where $\rho(0)=|\psi_0\rangle\langle\psi_0|$ is the 
density matrix of the system initial state.
We compute the MQC spectrum with $\hat{{\cal O}} = \hat{J}_z$. 
The decomposition of $\rho(t)$ in the eigenbasis of 
$\hat{J}_z$, $\{|m\ra\}_{m=-j}^{m=j}$, is
\be 
  \rho(t)=\sum_\ell\rho_\ell(t)=\sum_{\ell = -2j}^{2j} \sum_m\rho_{m+\ell,m}(t)|m+\ell\rangle\langle m|~,
\ee
where the $m$ index in the sum is such that $|m + \ell|\le j$ and
\begin{align} \label{Drho}
  \rho_{m+\ell,m}(t)&=\langle m+\ell|\rho(t)|m\rangle  \notag \\
    &=\langle m+\ell|e^{-i\hat H_1t}|\psi_0\rangle\langle\psi_0|e^{i\hat H_1t}|m\rangle \notag \\
    &=\zeta_{m+\ell}(t)\zeta^\ast_m(t),
\end{align}
with $\zeta_{m}(t)=\langle m|\psi_0(t)\rangle$ for $m = -j, \ldots, j$. Introducing the closure relation for the $\hat{H}_1$ eigenstates $\left\{|\phi_k^\alpha\rangle\right\}$, one can easily prove that
$\zeta_{m}(t)=\sum_{k,\alpha}\langle m|\phi_k^\alpha\rangle\langle \phi_k^\alpha|\psi_0\rangle e^{-iE_k t}$.
The eigenstates are such that $\hat H_1|\phi_k^\alpha\rangle=E_k|\phi_k^\alpha\rangle$ and the  $\alpha$ index is only needed in case there are degenerate states.

The next step is to assess whether the MQC spectrum is a useful tool ESQPT characterization. With this aim in mind,
we plot the time dependence of the $\zeta_m(t)$ factors in Figs.~\ref{QCspectrum}(a)-(c), the population time evolution, $I_0(t)$, in Figs.~\ref{QCspectrum}(d)-(f), and the dependence of $I_\ell(t)$ with  $\ell$ for selected time values in Figs.~\ref{QCspectrum}(g)-(i). In all cases, calculations are performed for a system with $\kappa=0.5$, $N=400$, and values of the quench parameter $\chi$ below, at, and above the critical value $\chi_c=-1/6$. 

As mentioned above, the time dependence of $\log|\zeta_m(t)|^2$ is shown as a heat map in Figs.~\ref{QCspectrum}(a-c), where we plot $\log{|\zeta_{m}(t)|}^2$ --
using the color scale introduced in panel~(a)-- versus the quantum number $n=m+j$ and time. 
Values of $|\zeta_m(t)|^2$ less than $10^{-12}$ have been truncated to zero. 
In a static problem, the transition state is known to be localized in 
the ground state of the Hamiltonian $\hat{H}(\kappa=0)$, i.e.~the $n = 0$ basis state~\cite{Santos2016}. 
In our case, $\zeta_{m}(t)$ is the time evolution of the $m_z$ component of $\ket{\psi_0}$. 
If the initial state remains in the broken-symmetry phase after the quench,  
the quenched state has large components for the 
the ground state and the first few excited states of $\hat{H}_1$.
As these eigenstates belong to the broken-symmetry phase, they have large components for basis states with large $n$ values. The resulting dynamics is characterized by regular oscillations of small amplitude in time as shown in Figs.~\ref{QCspectrum}(a, d) for $\chi=0.2\chi_c$.
The results obtained when the quenched state lies in the symmetric phase are similar to the previous ones, although the time evolution of $|\psi_0(t)\ra$
is now dependent on eigenstates of $\hat{H}_1$ 
with energies above the ESQPT critical energy, with an energy dependence that is more complex than in the symmetric case and with a wider distribution width in the $\hat J_z$ basis states. 
This leads to an evolution of $|\zeta_m(t)|^2$ with slightly larger amplitudes and 
more irregular oscillations for long times, 
as can be appreciated in Figs.~\ref{QCspectrum}(c, f) for $\chi=2\chi_c$.
Finally, if  $\chi=\chi_c$, the quench brings the system to the ESQPT critical energy. The results for this critical case are shown in  Figs.~\ref{QCspectrum}(b, e), where the 
$\ket{\psi_0(t)}$ state gets rapidly localized in the $n=0$ basis state, the ground state of $\hat{J}_z$.
The rapid localization can be explained from the divergence in the density of states, which makes the eigenstates with
energies close to the critical value dominate the system dynamics in the initial time.   
The highly localization feature of these eigenstates, originally explained in \cite{Santos2015, Santos2016}, can be explained in an intuitive manner considering that the critical ESQPT energy is associated with a unstable stationary point (a saddle point) in the classical limit, which favors the localization in the $n = 0$ state as the quantum system should have a large probability of localization in the origin. The initial localization observed in the evolution of $|\psi_0(t)\ra$ can be explained in this way, as well as the unstable dynamics at the critical energy of ESQPT,
with a fast spread of $\ket{\psi_0(t)}$ in the eigenstates of $\hat{H}_1$.
Therefore  the evolution is such that the time-evolved state is localized in more basis states than in the pre- or post-critical energy quench cases, even though $\ket{\psi_0(t)}$ remains  localized in the $n = 0$ basis state, an apparent conundrum already discussed in \cite{Chavez2023}. The unstable dynamics when the quenched state straddles the critical ESQPT energy is consistent with the results obtained in Refs.~\cite{WangP2019b, Cameo2020, Xu2020, Gamito2022II} and supports our initial assumption of considering the MQC spectrum and its width a valid probe for ESQPTs. We carry out detailed comparison of the dynamical features of the MQC spectrum width and FOTOC  in the following subsection.

Taking into consideration the MQC intensity definition in Eq.~(\ref{DfMQI}),  the $\ell$-th
MQC intensity of $\rho(t)$ is given by
\begin{align}\label{MQI}
  I_\ell[\rho(t)]=\mathrm{Tr}[\rho_\ell^\dag(t)\rho_\ell(t)]
  =\sum_m|\zeta_m(t)|^2|\zeta_{m+\ell}(t)|^2~.
\end{align}
Therefore, the MQC intensities, $\{I_\ell[\rho(t)]\}_{\ell=-2j}^{\ell=2j}$,  depend on $|\zeta_m(t)|^2$ for $m= -j \ldots j$. The maximal MQC intensity corresponds to the population intensity 
\be \label{ZeroM}
  I_0[\rho(t)]=\sum_m|\zeta_m(t)|^4~,
\ee
which is a time-dependent inverse participation ratio (IPR) that
measures the localization of 
the evolved state $|\psi(t)\rangle$ in the $\hat{J}_z$ basis. In Figs.~\ref{QCspectrum}(d-f), we show $I_0(t)$ for 
the same quench parameter values selected in Fig.~\ref{QCspectrum}(a-c). 
The depicted $I_0(t)$ results can be understood considering that $I_0(t)$ is an IPR and our previous discussion of the results obtained for $\log|\zeta_m(t)|^2$. 
In 
the $\chi < \chi_c$ case, from the results in Fig.~\ref{QCspectrum}(d), it is clear that the system remains unlocalized, with an IPR that undergoes small amplitude oscillations. As already mentioned, in this case the quenched $\ket{\psi_0(t)}$ depends on the first eigenstates of $\hat{H}_1$, which are not localized in the $\hat{J}_z$ basis states and have regular energy differences between them. 
For $\chi > \chi_c$ the quench brings the system to higher energies, into the broken-symmetry phase of the ESQPT. In this case the system dynamics is controlled by higher energy eigenstates of $\hat{H}_1$ and, for short times, eigenstates with larger energy differences modify
the system dynamics, leading t significant 
oscillations at times $t<100$ in Fig.~\ref{QCspectrum}(f). 
For longer times, more eigenstates 
contribute to the evolution of $\ket{\psi_0(t)}$. The energies associated to these eigenstates have a more complex spectrum than in the previous case, which explains the $I_0(t)$ irregular oscillations. 
The critical case, when $\chi = \chi_c$ and shown in  Fig.~\ref{QCspectrum}(e), is of special interest. 
The state of the system at the critical energy of ESQPT is localized but also dynamically unstable, making the system undergo a fast and strong localization to the $n=0$ 
basis state at short times, as already evinced in Fig.~\ref{QCspectrum}(b), 
followed by a fast delocalization with strongly irregular oscillations at low $I_0(t)$ values. 
To further explore the ESQPT influence on the MQC spectrum, 
 we have selected several time values, marked with vertical 
 lines in Figs.~\ref{QCspectrum}(d-f): $t=0$ (red full line), 
 $8$ (blue dashed line), $16$ (green dotted line), and $100$ (purple dash-dotted line). 
 We plot in Figs.~\ref{QCspectrum}(g-i) $I_\ell[\rho(t)]$ 
 as a function of $\ell$  for the selected time values, using the same colors and line styles. 
 The localization that occurs for $\chi = \chi_c$ is evinced 
 in the sharp maximum for time $t=8$ in Fig.~\ref{QCspectrum}(h). 
 Another interesting feature is the difference between  
 $I_\ell[\rho(t)]$ in Figs.~\ref{QCspectrum}(g) 
 and~\ref{QCspectrum}(i) for $t = 100$, that 
 can be traced back to the random oscillations in the latter case. 
Furthermore, Fig.~\ref{QCspectrum}(b) suggests that the maximum  number of $m_z$ values contributing to $|\zeta_m(t)|^2$ occurs at $\chi=\chi_c$,  
after the fast decay at initial times with a strong localization at $m_z = 0$. 
Therefore, at long times, the evolved MQC 
spectrum has a maximal width for $\chi \simeq \chi_c$.

To make clear how to characterize ESPQTs using $I_0[\rho(t)]$, we consider 
the maximum value of this quantity in a time interval $[0,\tau]$, denoted by $I_0^{max}$
\be \label{MaxI}
  I_0^{max}=\mathop{\mathrm{max}}\limits_{t\in[0,\tau]}\{I_0[\rho(t)]\}~,
\ee
\noindent where $\tau$ denotes a typical timescale for the short-term evolution of $I_0[\rho(t)]$.
In the present calculation we set $\tau=30$ and a careful check 
on cases with $\tau>30$ indicates that our conclusions are still valid.
In Fig.~\ref{Zero}(a), we plot the variation of $I_0^{max}$ with $\chi/\chi_c$ 
for two different system sizes ($N = 400, 800$) for a Hamiltonian with $\kappa=0.5$.
We clearly see that $I_0^{max}$ undergoes an abrupt change 
in the neighborhood of the critical quench,
independently of the system size $N$.
As a consequence of this, we suggest to use $I_0^{max}$ 
to check for the existence of an ESQPT in 
many-body quantum systems.
This is further confirmed by the results shown in Fig.~\ref{Zero}(b), 
where $I_0^{\max}$ is plot in a heat map as a function of 
$0\le\chi/\chi_c\le2 $ and $1.1\le\kappa/\kappa_c\le 1.9$ for a system size $N = 800$ and where 
the boundary between the system's phases is clearly 
located at $\chi/\chi_c\approx 1$, in good agreement with analytical results. 
 
 \begin{figure}
    \includegraphics[width=\columnwidth]{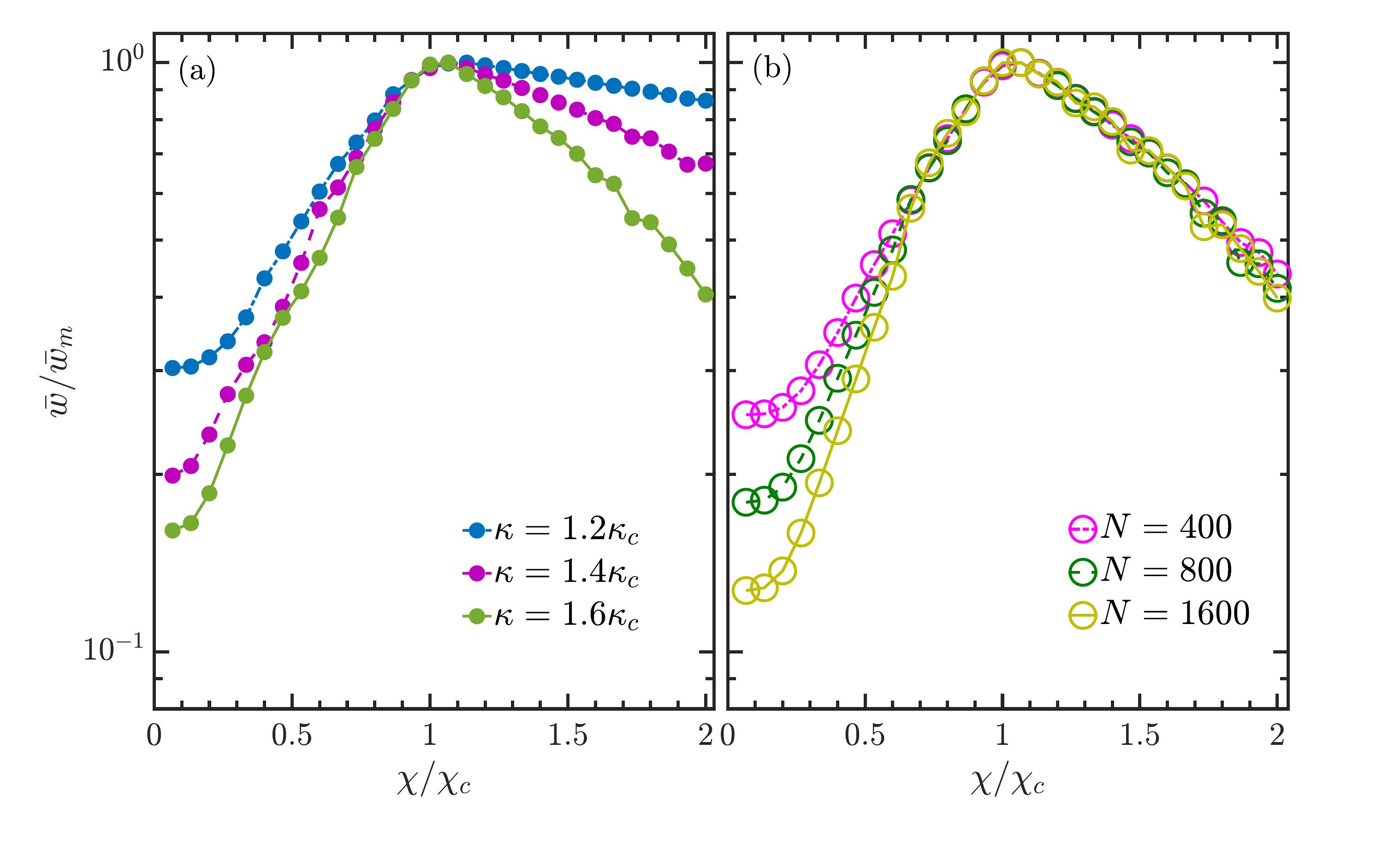}
    \caption{Normalized long-time averaged 
    MQC spectrum width, $\bar{w}/\bar{w}_m$ where $\bar{w}_m$ is the maximum value of $\bar{w}$.
    as a function of $\chi/\chi_c$ for
    (a) several values of the $\kappa$ control parameter with a system size $N=1000$ 
    and (b) different system size values, $N$, with $\kappa=1.6\kappa_c$.
    The maximum in $\bar{w}/\bar{w}_m$ marks the presence of the ESQPT.
    Other parameter values: $\kappa_c=1/3$ and $\chi_c$ is given by Eq.~(\ref{CrQ}).
    Axes in all figures are dimensionless.} 
    \label{Avgw}  
\end{figure}
 
 \subsection{The MQC spectrum width} \label{ForSb}
 
 The different amplitudes of $\zeta(t)$ shown in Figs.~\ref{QCspectrum}(a)-(c) 
 support our initial assumption that the MQC spectrum width could be a valid ESQPT probe.
 As we mentioned above, the MQC spectrum can be considered as an OTOC, and it is trivial 
 to realize that its width --- basically the second moment of 
 the MQC distribution --- is also a four-point correlator \footnote{Either prove this in an appendix or make a hint on how to reach this conclusion in a note.}.

 The MQC spectrum is symmetrically distributed around $\ell=0$ and its time-dependent width is
 given by the second moment in Eq.~(\ref{WMQCs}) for 
 $\rho(t)$ and $\hat{\mathcal{O}}=\hat{J}_z$,
 \be\label{Dwidth}
    w(t)=\Sigma[\rho(t),\hat{J}_z]=\sqrt{\sum_\ell\ell^2 I_\ell[\rho(t)]}~, 
 \ee
\noindent that is equal to the quantum Fisher information for pure quantum states~\cite{Garttner2018}. We now show that qualitative and quantitative features of $w(t)$ 
 can be used as ESQPT checks.
 
 \begin{figure}
    \includegraphics[width=\columnwidth]{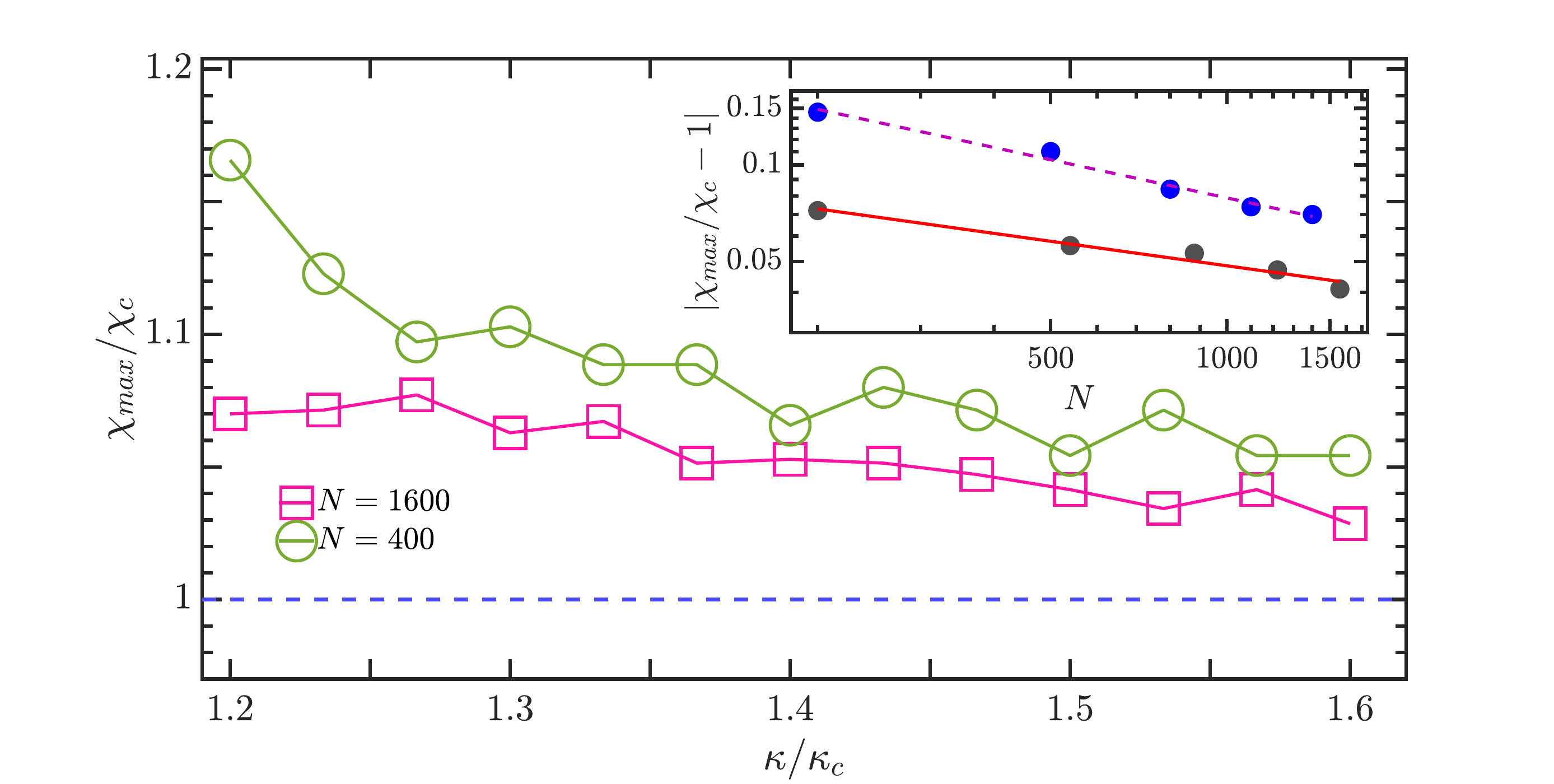}
    \caption{Normalized quench parameter  $\chi_{max}/\chi_c$ where $\chi_{max}$  is the quench parameter value for a maximum long-time 
    averaged MQC spectrum width, $\bar \omega_m$,
    as a function of $\kappa/\kappa_c$ for
    several size values $N=400$ and $1600$.
    The horizontal dashed line marks the $\chi_{max}/\chi_c=1$ value.
    The inset shows $|\chi_{max}/\chi_c-1|$ as a function 
    of $N$ on a log-log scale for $\kappa_1=1.3\kappa_c$ (blue dots)
    and $\kappa_1=1.5\kappa_c$ (gray dots).
    The purple dashed and red solid lines are of the form 
    $|\chi_{max}/\chi_c-1|=C_\kappa N^{-\beta_\kappa}$,
    with $C_{\kappa_1}=0.3952, \beta_{\kappa_1}=0.1902$ and
    $C_{\kappa_2}=0.2540, \beta_{\kappa_2}=1.273$.
    Other parameter values: $\kappa_c=1/3$ and $\chi_c$ is obtained from Eq.~(\ref{CrQ}).
    Quantities in all axes are dimensionless.} 
    \label{criticalp}  
\end{figure}
 
 Let us first consider how the dynamics of $w(t)$ is influenced by the ESQPT.  
 This is illustrated in Fig.~\ref{Timewidth}, where the time evolution of $w(t)$ for 
 different $\chi/\chi_c$ values is shown using log-log axes. 
 Vertical lines correspond to the particular times selected in Figs.~\ref{QCspectrum}(d-f). 
 For $\chi=0.2\chi_c$, before the critical quench,  
 the width oscillates in time around its initial value (solid dark orange line). 
 However, when the quench reaches its critical value for  $\chi=\chi_c$ (dashed light orange line), 
 we can distinguish four different regimes in the time evolution of the width $w(t)$ that are similar to the ones that have been observed with a  FOTOC in chaotic models~\cite{Chaves2019}: 
 (i) an approximately constant value  until $t\approx 1$, (ii) a decay to its minimum value, (iii) an exponential growth, 
 and (iv) $w(t)$ reaches its saturation value, oscillating around it in an irregular way. 
 The LMG is an integrable model, but the unstable stationary point 
 in the LMG Hamiltonian, marked with a red star in Fig.~\ref{Egsurface}, 
 makes the OTOC behave as expected for chaotic systems 
 whenever the system straddles the ESQPT critical energy~\cite{Cameo2020, Xu2020}. 
 Beyond the critical value of the quench, for $\chi>\chi_c$, 
 the system again oscillates regularly until it reaches its saturation value, as shown for $\chi = 2\chi_c$ with the dotted light orange lin in Fig.~\ref{Timewidth}. 
 In the inset, we show the long-time dependence using 
 lin-lin scale and we verify that the highest saturation value 
 is attained for the critical quench, the one that also 
 exhibits the highest dispersion in Fig.~\ref{QCspectrum}(b).

 To further confirm that $w(t)$ acts as a valid ESQPT probe, we consider
 the long-time averaged width, defined as
 \be \label{AvgWt}
   \bar{w}\equiv\lim_{T\to\infty}\frac{1}{T}\int_{t_0}^{t_0+T}w(t)dt~.
 \ee
 We chose $t_0$ values such that $t_0\gg1$.
 In the present calculations, we fixed $t_0=10^4$ and $T=10^3$ and
 we carefully checked that the obtained results are the same for larger 
 $t_0$ and $T$ values.
 
  \begin{figure*}
    \includegraphics[width=\textwidth]{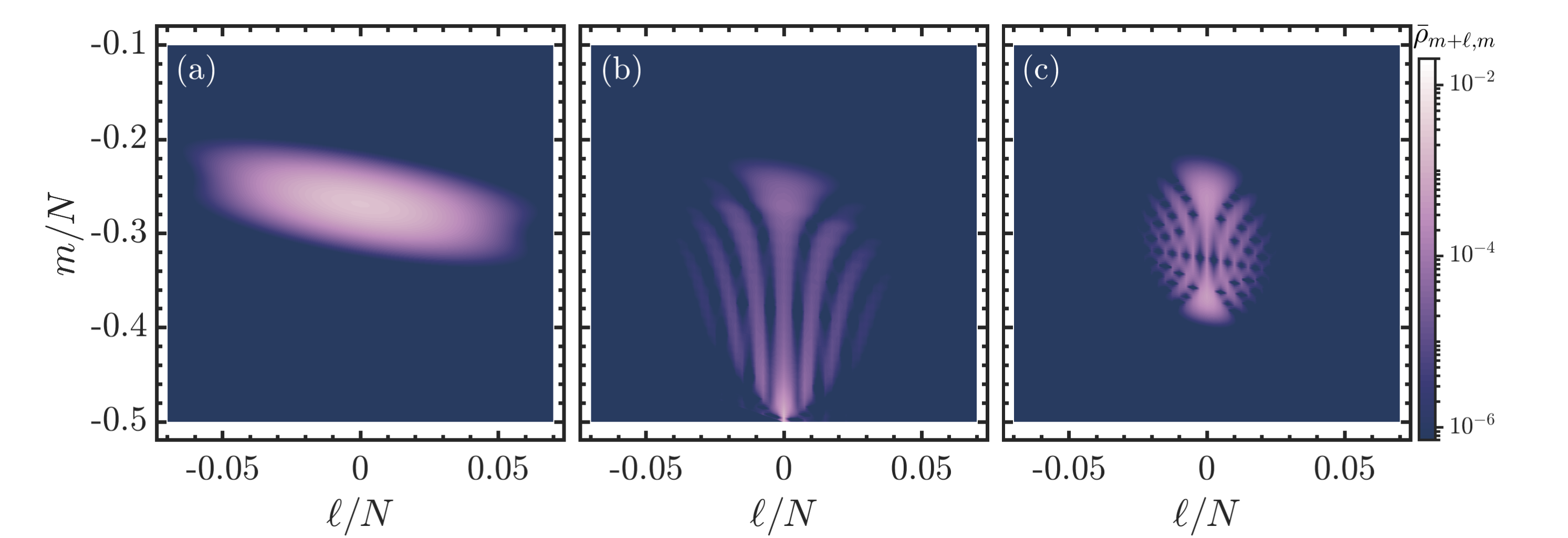}
    \caption{Heat map of  $\bar{\rho}_{m+\ell,m}$ in Eq.~(\ref{Rhom}) 
    as a function of $\ell/N$ and $m/N$ for
    (a) $\chi/\chi_c=0.2$, (b) $\chi/\chi_c=1$, and (c) $\chi/\chi_c=2$. 
    In the three cases $N=800$ and 
    $\kappa=1.5\kappa_c$ with $\kappa_c=1/3$, while
    the value of $\chi_c$ is given by Eq.~(\ref{CrQ}).
    The quantities depicted in the three panels are dimensionless.}
    \label{Cdml}  
\end{figure*}

Considering the results in Fig.~\ref{Timewidth},  it is expected that $\bar{w}$ should reach its
 maximum, $\bar{w}_m$,  at the vicinity of the critical quench.
 This is made clear with the scaled quantity $\bar{w}/\bar{w}_m$,  
 that is plotted as a function of $0\le\chi/\chi_c\le2$ in Fig.~\ref{Avgw}(a) 
 for different values of $\kappa$ and a fixed system size $N= 1000$, and in  Fig.~\ref{Avgw}(b) 
 for several $N$ values and a fixed $\kappa = 1.6/3$.
 One can clearly appreciate that, in both cases, $\bar{w}$ has a peak around $\chi/\chi_c=1$, regardless of 
 the strength of the control parameter $\kappa$ and the system size $N$, which
 means that $\bar{w}$ is an ESQPT precursor and the location of the peak
 provides a numerical estimation of the critical $\chi$ value: the peak location tends towards $\chi/\chi_c=1$ for
 increasing system size values. Moreover, the variation of $\bar{w}/\bar{w}_m$
around $\chi/\chi_c = 1$ is independent of the control parameter or the system size value, denoting the universal character of the scaling around the critical point. This effect is particularly conspicuous   in Fig.~\ref{Avgw}(b) where the values before and after the critical point are qualitatively different. This difference can be traced back to a scaling of $\bar{w}$ with $N$ for  $\chi/\chi_c < 1$ that switches to a $N^2$ scaling as the system gets close to the critical point and all the way to  $\chi/\chi_c = 2$. 
Such critical scaling of $\bar{w}/\bar{w}_m$ deserves further analysis for  a detailed understanding of its underlying  mechanism.
 
 The previous results are confirmed in Fig.~\ref{criticalp}, where the $\chi$ values that corresponds to
 maximum $\bar{w}$, denoted as $\chi_{max}$, are plotted as a function of $\kappa$ for $N = 400$ and $1600$.
 To better grasp how $\chi_{max}$ tends to $\chi_c$ as $N$ increases, we
 include an inset in Fig.~\ref{criticalp} where the variation of
 $|\chi_{\max}/\chi_c-1|$ with the system size $N$ for $\kappa=1.3/3$ 
 and $\kappa=0.5$ are plotted using a log-log scale.
 It can be easily appreciated that, regardless of the $\kappa$ value, the scaling behavior of 
 $|\chi_{max}/\chi_c-1|$ is well fitted by a power law,
 $|\chi_{max}/\chi_c-1|=C_\kappa N^{-\beta_\kappa}$,
 where $C_\kappa$ and $\beta_\kappa$ are $\kappa$ dependent quantities. Hence, $\chi_{max}\to\chi_c$ as $N$ goes to infinite.

 \subsection{MQC spectrum of the long-time averaged state} \label{ForSc}
 
We now discuss how to unveil ESQPT signatures through the MQC spectrum
 of the long-time averaged state, which is defined as
 \begin{align} \label{eq:Avgrho}
   \bar{\rho}&\equiv\lim_{T\to\infty}\frac{1}{T}\int_0^Tdt\,\rho(t) \notag \\
    &=\lim_{T\to\infty}\frac{1}{T}\int_0^Tdt\, e^{-i\hat{H}_1t}
             |\psi_0\ra\la\psi_0|e^{i\hat{H}_1t} \notag \\
    &=\sum_k|\la\phi_k|\psi_0\ra|^2|\phi_k\ra\la\phi_k|~.
 \end{align}
 In the last step we have used the closure relation
 $\hat{\mathbb{I}}=\sum_k|\phi_k\ra\la\phi_k|$ with $|\phi_k\ra$ being the
 $k$th eigenstates of $\hat{H}_1$.
 Moreover, the integration has been
 carried out considering that there is no degeneracy in the $\hat{H}_1$ energy spectrum. 
 This assumption is valid because we start from a positive parity state, 
 and the operator $\hat{J}_z$ does not mix different parity states. 
 It is worth noting that $\bar{\rho}$ is a mixed state.
 
 To calculate the MQC spectrum of $\bar{\rho}$, 
 we first express this state in the  $\hat{J}_z$ basis as
 \be
   \bar{\rho}=\sum_\ell\bar{\rho}_\ell
     =\sum_\ell\sum_m\bar{\rho}_{m+\ell, m}|m+\ell\rangle\langle m|~,
 \ee
 where
 \begin{align}\label{Rhom}
   \bar{\rho}_{m+\ell,m}&=\la m+\ell|\bar{\rho}|m\ra \notag \\
   &=\sum_{k}\la m+\ell|\phi_k\ra\la \phi_k|m\ra|\la\psi_0|\phi_k\ra|^2~.
 \end{align}
Finally, the resulting $\ell$-th component of the MQC spectrum is
 \begin{align} \label{AvgrhoMQC}
   I_\ell(\bar{\rho})=\mathrm{Tr}[\bar{\rho}_\ell^\dag\bar{\rho}_\ell]
      =\sum_m\left|\bar{\rho}_{m+\ell,m}\right|^2~.
 \end{align}
 
 The latter formula indicates that the MQC spectrum of $\bar{\rho}$ 
 is determined by the dependence of
 $\bar{\rho}_{m+\ell,m}$ on $\ell$ and $m$. 
 This is made clear in Fig.~\ref{Cdml} panels, where $\bar{\rho}_{m+\ell,m}$ 
 is depicted as a heat map dependent on the system-size scaled values of $m$ and $\ell$ for three different 
 values of $\chi/\chi_c$ with $\kappa=0.5$. 
 In the three panels is evinced the variation of 
 $\bar{\rho}_{m+\ell,m}$ as the system goes through the critical quench. 
 In particular, in the $\chi/\chi_c<1$ case, $\bar{\rho}_{m+\ell,m}$ is a smooth function 
 deprived of nodes and that extends in a wide $\ell$ range, 
  as shown in Fig.~\ref{Cdml}(a). 
 This is due to the large overlap between 
 $\ket{\phi_0}$ and $\ket{\psi_0}$.
 On the contrary, in the  $\chi/\chi_c>1$ phase, 
 there are several eigenstates $\ket{\phi_k}$ for large $k$ 
 values contributing to $\ket{\psi_0}$.
 Hence, the summation of their distributions in the basis states as in Eq.~(\ref{Rhom})
 is more complex as shown in Fig.~\ref{Cdml}(c), where 
 $\bar{\rho}_{\ell+m,m}$ oscillates in both variables due to interference effects
 and quickly decays to zero as outside a range of $|\ell|$ and $m$ values.
 As illustrated in Fig.~\ref{Cdml}(b), at the critical 
 value $\chi/\chi_c=1$,  $\bar{\rho}_{m+\ell,m}$ displays 
 a dependence on $\ell$ and $m$ that differs from the two cases previously described. 
 It again exhibits a complex, oscillating pattern as in previous case, but the oscillations  extend along large
 $m$ and $\ell$ ranges. I addition to this, the most salient feature is the 
 strong localization in the $n = 0$ state. The results for the critical case can be understood from the critical character of this state and the
 divergence of the density of states at the ESQPT critical energy, which
 makes $\ket{\psi_0}$ become a highly localized state in the $\hat J_z$ basis.

    \begin{figure*}
    \includegraphics[width=\textwidth]{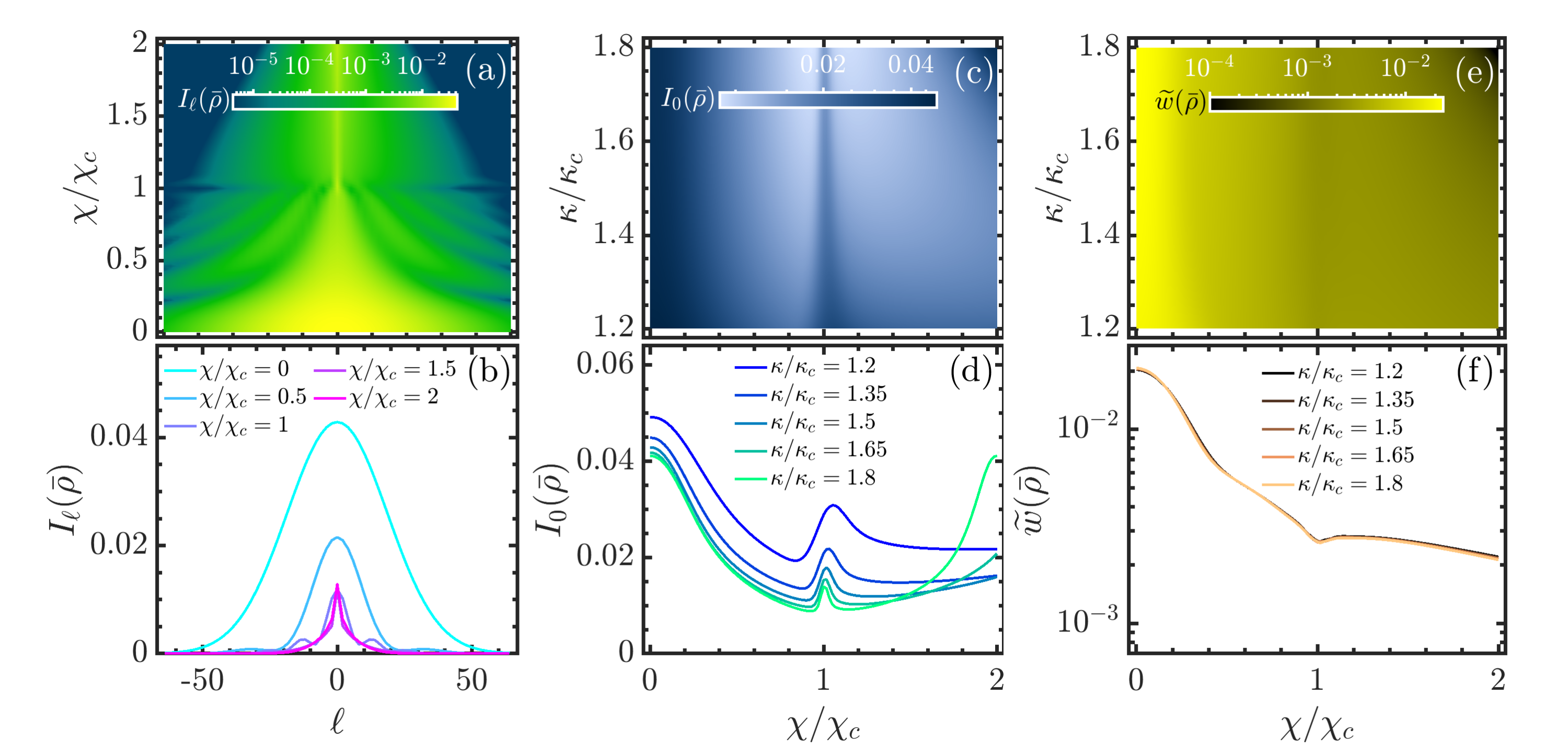}
    \caption{(a) Heat map of the long-time averaged state 
    MQC spectrum $I_\ell(\bar{\rho})$ defined in Eq.~(\ref{AvgrhoMQC})
    as a function of $\ell$ and $\chi/\chi_c$ for $\kappa=1.5\kappa_c$.
    (b) $I_\ell(\bar{\rho})$ as a function of $\ell$ for several values of 
    the quench parameter $\chi/\chi_c$ (see legend) with $\kappa=1.5\kappa_c$.
    The crossing of the ESQPT drives a significant change in the MQC spectrum.
    (c) Heat map of the zero component of the  long-time averaged state  MQC spectrum , $I_0(\bar{\rho})$ (see Eq.~(\ref{ZeroAvg})), as a function of $\chi/\chi_c$ and $\kappa/\kappa_c$.
    (d) $I_0(\bar{\rho})$ as a function of $\chi/\chi_c$ for 
    selected values of $\kappa/\kappa_c$ (see legend).
    (e) Heat map of the size-scaled width of the long-time averaged state 
    MQC spectrum, $\widetilde{w}(\bar{\rho})=w(\bar{\rho})/N$,
    as a function of $\chi/\chi_c$ and $\kappa/\kappa_c$.
    (f) $\widetilde{w}(\bar{\rho})$ as a function of $\chi/\chi_c$ for several values of 
    $\kappa/\kappa_c$ (see legend).
    In panels (c) to (f) the local maximum in $I_0(\bar{\rho})$ and the local minimum in $w(\bar{\rho})/N$
    make these two quantities possible ESQPT probes.
    In all panels $N=2j=800$, $\kappa_c=1/3$, and $\chi_c$ is obtained from Eq.~(\ref{CrQ}).
    The quantities depicted in the different axes of all figures are dimensionless.} 
    \label{Rhoag}  
\end{figure*}
 
 It is clear from the heat maps of $\bar{\rho}_{m+\ell,m}$ in Figs.~\ref{Cdml}(a-c) that 
 the model ESQPT shapes the MQC spectrum. 
 This is indeed confirmed in Fig.~\ref{Rhoag}(a), where we depict 
 the MQC spectrum $I_\ell(\bar{\rho})$ as a heat map depending on $\ell$ and $\chi/\chi_c$ 
 for the $\kappa=0.5$ case. In this panel it is clear how, for increasing 
 $\chi/\chi_c$ values, the  $\ell$ range associated with large $I_\ell(\bar{\rho})$
 exhibits a rapid decrease as $\chi$ passes through the critical value. 
 Moreover, the higher values of $\bar{\rho}_{m+\ell,m}$ for the $\chi/\chi_c<1$ case shown in Fig.~\ref{Cdml}(a) explain the large amplitudes of the MQC spectrum in Fig.~\ref{Rhoag}(a).
 This is clearly shown in Fig.~\ref{Rhoag}(b), where the MQC spectrum for several 
 $\chi/\chi_c$ values is plotted.
 
 The results in Figs.~\ref{Rhoag}(a) and \ref{Rhoag}(b) 
 confirm the $\bar{\rho}$ MQC spectrum  as a valid 
 candidate for an ESQPT indicator in many-body quantum system. 
 To further demonstrate the usefulness of the MQC 
 distribution of $\bar{\rho}$ for ESQPT characterization, we now focus on the
 $I_\ell(\bar{\rho})$ zero mode, denoted as $I_0(\bar{\rho})$ 
 \be \label{ZeroAvg}
    I_0(\bar{\rho})=\sum_m|\bar{\rho}_{m,m}|^2~,
 \ee 
 where $\bar{\rho}_{m,m}=\sum_ {k}|\la m|\phi_k\ra|^2|\la\psi_0|\phi_k\ra|^2$.
 
Fig.~\ref{Rhoag}(c) is a heat map for $I_0(\bar{\rho})$ as a function 
of the parameters $\kappa$ and $\chi$ where the value of $I_0(\bar{\rho})$ 
exhibits a local maximum at the  critical quench value, $\chi/\chi_c=1$.  
 This is even more evident in Fig.~\ref{Rhoag}(d), where the 
 dependence of $I_0(\bar{\rho})$ on 
 $\chi/\chi_c$ for several $\kappa$ values (see plot legend) is depicted. 
 Hence, the zero mode of the  $\bar{\rho}$ MQC spectrum is 
 sensitive to the presence of the ESQPT. The local maximum of  
 $I_0(\bar{\rho})$ at $\chi/\chi_c=1$ can be explained from  
 the strong localization in the $n=0$ basis state an d the large range of $m$ values contributing to  $I_0(\bar{\rho})$, as shown in Fig.~\ref{Cdml}(b).
 To further understand this behavior, we could consider a simpler system, 
 where the quench Hamiltonian was $\hat{H}(\kappa=0)\propto\hat{J}_z$. 
 In this case, the population of the MQC spectrum for a long-time 
 average state would coincide with the long-time average of the Loschmidt echo (See Refs.~\cite{WangH2017,Gamito2022II}). 
 This explains the  similarities between Fig.~\ref{Rhoag}(d) 
 and Fig.~5(d) of Ref.~\cite{WangH2017}, where Q.~Wang and H.~T.~Quan 
 found a comparable behavior for the long-time average of the Loschmidt echo. 
 With this setup, the higher the localization in the eigenbasis 
 of the quench Hamiltonian, the larger the value of the $I_0(\bar{\rho})$. 
 A similar phenomenology is observed in our system. 
 Looking at Fig.~\ref{Cdml}, we realize that $I_0(\bar{\rho})$ 
 in Fig.~\ref{Rhoag}(d) is higher when  the state is well localized in the quench eigenbasis.

 Furthermore, as the MQC spectrum has a different width in the two ESQPT phases,
  the $\bar{\rho}$ MQC spectrum width should also change due to the presence of the ESQPT. 
 To verify this point, we show in Fig.~\ref{Rhoag}(e) the scaled width  of the $\bar{\rho}$ MQC spectrum , $\tilde{w}(\bar{\rho})=w(\bar{\rho})/N$, 
 as a function of $\kappa$ and $\chi$.
 It is clear that $\tilde{w}$ has a kink at the critical point, regardless of the $\kappa$ value,
 in contrast to the case of long time averaged $w(t)$ which exhibits a maximum in the neighborhood 
 of the critical point, as seen in Fig.~\ref{Avgw}.
 The kink in $w(\bar{\rho})/N$ is clearly appreciated in Fig.~\ref{Rhoag}(f), where this quantity is depicted as a function of
 $\chi/\chi_c$ for several $\kappa$ values, and it can be traced back to the
 low $\bar{\rho}_{m+\ell,m}$ values at the critical point ---cf.~Fig.~\ref{Cdml}(b).
 Finally, it is important to appreciate that the evolution of $w(\bar{\rho})/N$ as a function of $\chi/\chi_c$ is basically
independent of the $\kappa$ value.

\section{Conclusions}\label{FivS}

In the present article, we have investigated, through the MQC spectrum. the influence on quantum coherence of the LMG model ESQPT, a logarithmic divergence in the mean-field limit of the density of states.  By using a sudden-quench protocol and
expanding the evolved state in the $\hat{J}_z$ quasispin basis, 
we have found that the underlying ESQPT has a strong
impact on the time evolution of the MQC spectrum, whose temporal dependence at the quench critical parameter value is a reliable indicator of the presence of an ESQPT. The time evolution of the MQC spectrum can be understood by studying  the $\zeta(t)$ coefficients defined in Eq.~\eqref{Drho}. Particularly, the system gets quickly localized in the $\hat{J}_z$ basis state with a minimum value of $m_z$ at the critical value of the quench parameter. We have also shown that this localization leads into a clear maximum at short times in the zero-mode MCQ spectrum, $I_{\ell=0}$ followed by a high dispersion (see~Fig.~\ref{QCspectrum}e). 

A high dispersion is also observed when we study the MQC spectrum versus the differences $\ell$. To understand this behavior, we plotted the width of the distribution and identified different ways followed by the system to reach saturation. In particular, in the quench to the critical energy of the ESQPT case, we observe regimes previously identified in other OTOCs and a maximal saturation value. Another result of interest in the critical quench is that the long-time saturation width peaks at locations that tend to the critical point for increasing system sizes. 

We have also analyzed the properties of the MQC distribution for the long-time
averaged state, which is defined as the long-time average of the evolved state [cf.~Eq.~(\ref{eq:Avgrho})] and we have observed that the ESQPT is resposible for an abrupt change in the MQC
spectrum of the long-time averaged state.
Moreover, its zero mode and width also stand as good detectors of the ESQPT in 
the LMG model. The first one is maximum and the second minimum when the quench parameter reaches its critical value.

The present work focus on the LMG model but our conclusions are a consequence of the nature of the  
ESQPT in this model, characterized by a logarithmic divergence of the density of states at certain excited energy. As explained, despite its simplicity, the LMG model has proved a suitable platform to study GSQPTs. ESQPTs and other critical phenomena. Furthermore, there are several experimental realizations of this model and the connection between the MQC spectrum and a FOTOC [cf.~Eq.~(\ref{MQCFotoc})]    allows for the experimental access to the MQC spectrum through protocols 
that could be implemented in different platforms \cite{Garttner2017,Swan2020} and do not 
require full quantum state tomography at each time. 
Based on these facts, we expect that our results may
motivate further experimental efforts to investigate ESQPTs using the MQC spectrum. In fact, ESQPTs akin to the one in the LMG model appear in many other models with a single effective degree of freedom and, therefore, our findings will also apply for other systems that undergo the same kind of 
ESQPT as the LMG model.
A natural extension of the present work is a systematic exploration of 
the ESQPT effects on the MQC spectrum in other quantum many-body systems.

\acknowledgments

Q.~W.~acknowledges support from the 
the Slovenian Research and Innovation Agency (ARIS) under the Grant
No.~J1-4387 and P1-0306, 
National Science Foundation of China under
Grant No.~11805165, Zhejiang Provincial Nature Science Foundation under Grant
No.~LY20A050001. 
This project has also received funding from the 
Grant PID2022-136228NB-C21 funded by MICIU/AEI/
10.13039/501100011033 and, as appropriate, by “ERDF A way of making Europe,” by
“ERDF/EU,” by the “European Union,” or by the “European Union
NextGenerationEU/PRTR.” and by the FEDER-UHU project POSH-AI, EPIT1462023. 
JKR also acknowledges support from a 
Spanish Ministerio de Universidades \textit{Margarita Salas} Fellowship.
Computing resources supporting this work were partly provided by the CEAFMC and
Universidad de Huelva High Performance Computer (HPC@UHU) located in the Campus
Universitario el Carmen and funded by FEDER/MINECO project UNHU-15CE-2848.

\bibliographystyle{apsrev4-2}

%

\end{document}